\documentclass[12pt]{revtex4}
\usepackage{graphicx}
\usepackage{amsmath}

\begin{document}

\title{Ab initio study of alanine polypeptide chains twisting}

\author{Ilia A. Solov'yov\footnote{On leave from the A.F. Ioffe Institute, St. Petersburg, Russia.
E-mail: ilia@th.physik.uni-frankfurt.de}, Alexander V. Yakubovitch,
        Andrey V. Solov'yov\footnote{On leave from the A.F. Ioffe Institute, St. Petersburg, Russia.
E-mail: solovyov@fias.uni-frankfurt.de},
        and Walter Greiner}

\address{Frankfurt Institute for Advanced Studies,
Max von Laue Str. 1, 60438 Frankfurt am Main, Germany}

\begin{abstract}
We have investigated the potential energy surfaces for alanine
chains consisting of three and six amino acids. For these molecules
we have calculated potential energy surfaces as a function of the
Ramachandran angles $\varphi$ and $\psi$, which are widely used for
the characterization of the polypeptide chains. These particular
degrees of freedom are essential for the characterization of
proteins folding process. Calculations have been carried out within
{\it ab initio} theoretical framework based on the density
functional theory and accounting for all the electrons in the
system. We have determined stable conformations and calculated the
energy barriers for transitions between them. Using a thermodynamic
approach, we have estimated the times of characteristic transitions
between these conformations. The results of our calculations have
been compared with those obtained by other theoretical methods and
with the available experimental data extracted from the Protein Data
Base. This comparison demonstrates a reasonable correspondence of
the most prominent minima on the calculated potential energy
surfaces to the experimentally measured angles $\varphi$ and $\psi$
for alanine chains appearing in native proteins. We have also
investigated the influence of the secondary structure of polypeptide
chains on the formation of the potential energy landscape. This
analysis has been performed for the sheet and the helix
conformations of chains of six amino acids.
\end{abstract}

\maketitle

\section{Introduction}

It is well known that proteins consist of amino acids whose number
may vary in the range from hundreds up to tens of thousands. Small
fragments of proteins are usually called polypeptide chains or
polypeptides. This work is devoted to a study of the conformational
properties of alanine polypeptide chains.

Since recently, it became possible to study experimentally small
fragments of proteins and polypeptides in the gas phase with the use
of the MALDI mass spectroscopy \cite{Karas88,Karas00,Karas03,Wind04}
and the ESI mass spectroscopy \cite{Fenn89,Hvelplund04}. From
theoretical viewpoint, investigation of small polypeptides is of
significant interest because they can be treated by means of {\it ab
initio} methods which allowing accurate comparison of theoretical
predictions with experiment. The results of {\it ab initio}
calculations can be then utilized for the development of model
approaches applicable for the description of larger and more complex
protein structures.

Polypeptides are characterized by the primary and the secondary
structure \cite{Ptizin_book,Muelberg_book,Protbase,Rubin04}.
Different geometrical configurations of a polypeptide are often
called as the conformations. One can expect that chemical and
physical properties of various conformations of complex molecules
might differ significantly. The number of various conformations
(isomeric states) grows rapidly with the growth of a system size.
Thus, a search for the most stable conformations becomes an
increasingly difficult problem for large molecules. With the help of
the NMR spectroscopy and the X-rays diffraction analysis it has been
shown \cite{Protbase}, that the sheet and the helix structures are
the most prominent elements of the protein secondary structure.

The main difference between the sheet and the helix structures is
due to the difference of the dihedral angles formed by the atoms of
the polypeptide chains in the two cases. These degrees of freedom
are responsible for the transition of the molecule from one
conformation to another. By increasing the temperature of the
system, the degrees of freedom responsible for twisting of the
polypeptide chain can be activated. The study of this transition and
evaluation of its characteristic duration are of significant
interest, because this problem is closely related to one of the most
intriguing problems of the protein physics - the protein folding. To
study this transition it is necessary to investigate the potential
energy surface of amino acid chains with respect to their twisting.
Besides the protein folding, the potential energy landscapes of
polypeptides carry a lot of detail and useful information about the
structure of these molecules.

In the present paper we have studied the potential energy surfaces
for small alanine chains. These molecules were chosen because they
are often present in native proteins as fragments, and also because
they allow {\it ab initio} theoretical treatment due to their
relatively small size.

Previously, only glycine and alanine dipeptides were studied in
detail. Sometimes their analogues were used to reduce the
computational costs (for example,
(S)-$\alpha$-(formylamino)propanamide). In refs.
\cite{Head-Gordon91,Gould94,Zhi-Xiang04} alanine and glycine
dipeptides were investigated within the Hartree-Fock theory. In
these papers the potential energy surfaces were calculated versus
the twisting angles of the molecules. Different stable states of the
dipeptides, corresponding to different molecular conformations, were
determined. Each stable state of the molecule was additionally
studied on the basis of the perturbation theory, which takes into
account many-electron correlations in the system. In refs.
\cite{Percel03,Hudaky04,Improta04,Vargas02,Kashner98,Salahub01}
different conformations and their energies were determined within
the framework of the density functional theory. In ref.
\cite{Salahub01} dynamics of the alanine dipeptide analog was
discussed and the time of the transitions between the two
conformations of the alanine dipeptide was found.

A number of papers were devoted to the study of tripeptides. In
refs. \cite{Woutersen01_1,Woutersen02,Stock02,Stock03,Stock03_1}
dynamics of the alanine and glycine tripeptides was studied by means
of classical molecular dynamics and with the use of semi-empirical
potentials (such as GROMOS, CHARMM and AMBER). In \cite{Torii98}
within the framework of the Hartree-Fock theory several stable
conformations of alanine and glycine tripeptides were found. In ref.
\cite{Stenner01} the Raman and IR spectra for alanine and glycine
tripeptides were measured in neutral, acidy and alkali environments.

Polypeptides have been studied less. We are aware of only several
related papers. In particular, stable conformations of neutral and
charged alanine hexapeptides were obtained with the use of empirical
potentials and discussed in ref. \cite{Levy}. Experimental NMR study
of various conformations of alanine heptapeptides at different
temperatures was carried out in ref. \cite{Shi02}. In ref.
\cite{Garcia03} with the use of empirical molecular dynamics based
on Monte-Carlo methods, a polypeptide consisting of 21 amino acids
was described.

In the present paper we have performed an {\it ab initio}
calculation of the multidimensional potential energy surface for the
alanine polypeptide chains consisting of three and six amino acids.
The potential energy surface versus twisting degrees of freedom of
the polypeptide chain has been calculated. The calculations have
been performed within {\it ab initio} theoretical framework based on
the density functional theory (DFT) accounting for all the electrons
in the system. Previously, this kind of calculations were performed
only for dipeptides (see, e.g.,
\cite{Head-Gordon91,Gould94,Salahub01}). For larger molecules, only
a few conformations were considered (see citations above). We have
calculated the energy barriers for the transitions between different
conformations and determined the energetically most favorable ones.
Using a thermodynamic approach, we have estimated times of the
characteristic transitions between the most energetically favorable
conformations. The results of our calculation have been compared
with other theoretical simulations and with the available
experimental data. We have also analyzed how the secondary structure
of polypeptide chains influences the potential energy landscapes. In
particular, the role of the secondary structure in the formation of
stable conformations of the chains of six amino acids being in the
sheet and in the helix conformations has been elucidated. Some
preliminary results of our work were published as electronic
preprints \cite{twisting_preprint,fission_preprint}.

Our paper is organized as follows.  In section \ref{theory} we give
a brief overview of theoretical methods used in our work. In section
\ref{results} we present and discuss the results of our
computations. In section \ref{conclusions} we draw a conclusion to
this paper. The atomic system of units, $|e|=m_e=\hbar=1$, is used
throughout the paper unless other units are indicated.

\section{Theoretical methods}
\label{theory}

In the present paper we study the multidimensional potential energy
surfaces for alanine polypeptides within the framework of the
density functional theory. The potential energy surfaces are
multidimensional functions of atomic coordinates. In our work the
potential energy surfaces are considered as a function of the
dihedral angles formed by the atoms of the polypeptide chain. For
this calculation the Born-Oppenheimer approximation allowing to
separate the motion of the electronic and ionic subsystems is used.

The density functional theory (DFT) is a common tool for the
calculating of properties of quantum many body systems in which many
electron correlations play an important role. The DFT formalism is
well known and can be found in many textbooks (see e.g.
\cite{LesHouches,ISACC2003}). Therefore in our work we present only
the basic equations and ideas of this method.

Electronic wave functions and energy levels within the framework of
DFT are obtained from the Kohn-Sham equations, which read as (see
e.g. \cite{LesHouches,ISACC2003}):

\begin{equation}
\left( \frac{\hat p^2}2+U_{ions}+V_{H}+V_{xc}\right) \psi_i
=\varepsilon _i \psi _i,\ \ \ \ \ \ i=1...N
\end{equation}

\noindent where the first term represents the kinetic energy of the
$i$-th electron with the wavefunction $\psi_i$ and the energy
$\varepsilon_i$, $U_{ions}$ describes the electron attraction to the
ionic centers, $V_{H}$ is the Hartree part of the interelectronic
interaction ref. \cite{Lindgren}, $V_{xc}$ is the local
exchange-correlation potential.

The exchange-correlation potential is defined as a functional
derivative of the exchange-correlation energy functional:

\begin{equation}
V_{xc}=\frac{\delta E_{xc}[\rho]}{\delta \rho(\vec r)}, \label{Vxc}
\end{equation}

\noindent Equation (\ref{Vxc}) is exact and follows from the
Hohenberg theory ref. \cite{Hohenberg64}. However, no unique
potential $E_{xc}$, universally applicable for different systems and
conditions, has been found so far.

Approximate functionals employed by the DFT usually partition the
exchange-correlation energy into two parts, referred to as the {\it
exchange} and the {\it correlation} terms:

\begin{equation}
E_{xc}[\rho]= E_x(\rho)+E_c(\rho) \label{ex_core}
\end{equation}

\noindent Both terms are the functionals of the electron density,
which can be of two distinctly different types: either a {\it local}
functionals depending only on the electron density $\rho$, or a {\it
gradient-corrected} functionals depending on both $\rho$ and its
gradient, ${\bf \nabla} \rho$. A variety of exchange correlation
functionals can be found in literature. In our work we have used the
hybrid Becke-type three-parameter exchange functional ref.
\cite{Becke88} paired with the gradient-corrected Lee, Yang and Parr
correlation functional ($B3LYP$) refs. \cite{LYP,Parr-book}.

\section{Results and Discussion}
\label{results}

\subsection{Determination of the polypeptide twisting degrees of freedom}

\begin{figure}[h]
\includegraphics[scale=0.55,clip]{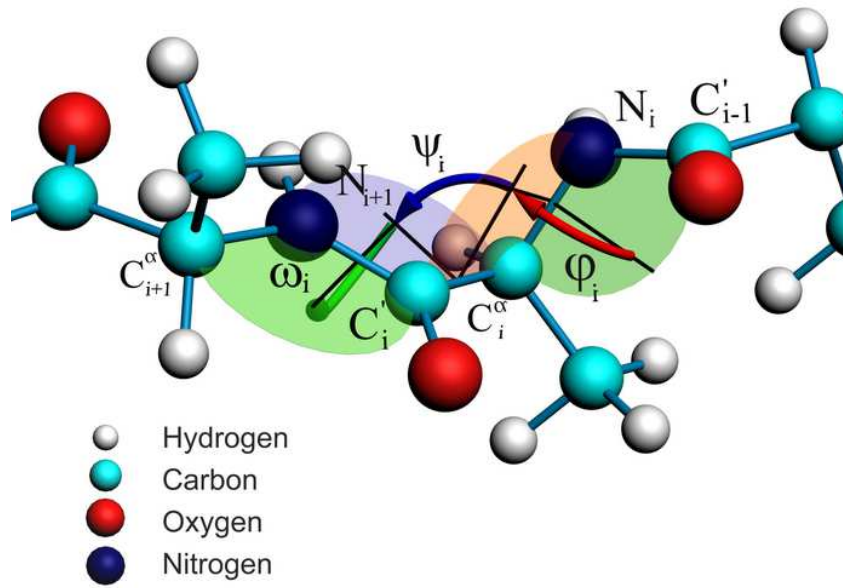}
\caption{Dihedral angles $\varphi$ and $\psi$ used to characterize
the potential energy surface of the polypeptide chain.}
\label{angles_def}
\end{figure}

In this section we present the potential energy surfaces for the
glycine polypeptide chains calculated versus dihedral angles
$\varphi$ and $\psi$ defined in figure \ref{angles_def}. In
particular, we focus on the chains consisting of three and six amino
acids.

Both angles are defined by the four neighboring atoms in the
polypeptide chain. The angle $\varphi_i$ is defined as the dihedral
angle between the planes formed by the atoms
($C_{i-1}^{'}-N_{i}-C_i^{\alpha}$) and
($N_{i}-C_i^{\alpha}-C_i^{'}$). The angle $\psi_i$ is defined as the
dihedral angle between the ($N_{i}-C_{i}^{\alpha}-C_i^{'}$) and
($C_{i}^{\alpha}-C_i^{'}-N_{i+1}$) planes. Beside the angles
$\varphi_i$ and $\psi_i$ there is an angle $\omega_i$, which is
defined as the dihedral angle between
($C_{i}^{\alpha}-C_i^{'}-N_{i+1}$) and
($C_{i}^{'}-N_{i+1}-C_{i+1}^{\alpha}$) planes. The atoms are
numbered from the $NH_2-$ terminal of the polypeptide. The angles
$\varphi_i$, $\psi_i$ and $\omega_i$ take all possible values within
the interval [$-180^{o}$;$180^{o}$]. For the unambiguous definition
we count the angles $\varphi_i$, $\psi_i$ and $\omega_i$ clockwise,
if one looks on the molecule from its $NH_2-$ terminal (see fig.
\ref{angles_def}). This way of angle counting is the most commonly
used \cite{Rubin04}.

The angles $\varphi_i$ and $\psi_i$ can be defined for any amino
acid in the chain, except the first and the last ones. Below we omit
the subscripts and consider angles $\varphi$ and $\psi$ for the
middle amino acid of the polypeptide.

\subsection{Optimized geometries of alanine polypeptides}

In order to study twisting of a polypeptide chain one needs first to
define its initial structure. Although, the number of its
conformations increases with the growth of the molecule size, there
are certain types of polypeptide structure, namely the sheet and the
helix conformations, which are the most typical. In the present
paper we have investigated twisting of the polypeptide chains of the
sheet and the helix conformations. By varying the angles $\varphi$
and $\psi$ in the central amino acid one can create the structure of
the polypeptide differing significantly from the pure sheet or helix
conformations. If the structure of a polypeptide can be transformed
to a helix or a sheet one by a trivial variation of $\varphi$ and
$\psi$, such polypeptides for the sake of simplicity are referred
below as belonging to the group of the helix or the sheet structure,
respectively.
\begin{figure}[h]
\includegraphics[scale=0.84,clip]{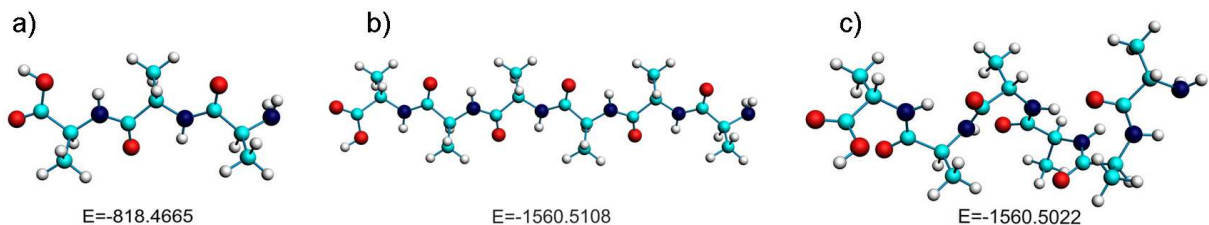}
\caption{Optimized geometries of alanine polypeptide chains
calculated by the B3LYP/6-31++G(d,p) method: a) Alanine tripeptide;
b) Alanine hexapeptide (sheet conformation); c) Alanine hexapeptide
(helix conformation).} \label{stable_geom}
\end{figure}

In figure \ref{stable_geom} we present the optimized geometries of
alanine polypeptide chains that have been used for the exploration
of the potential energy surfaces. All geometries were optimized with
the use of the B3LYP functional. Figure \ref{stable_geom}a shows the
alanine tripeptide structure. In the present work we choose the
sheet conformation, because the tripeptide is too short to form the
helix conformation. Figures \ref{stable_geom}b and
\ref{stable_geom}c show alanine hexapeptide in the sheet and the
helix conformations, respectively. The total energies (in atomic
units)of the molecules are given below the images.

\subsection{Polypeptide energy dependance on the dihedral angle $\omega$}

\begin{figure}[h]
\includegraphics[scale=0.9,clip]{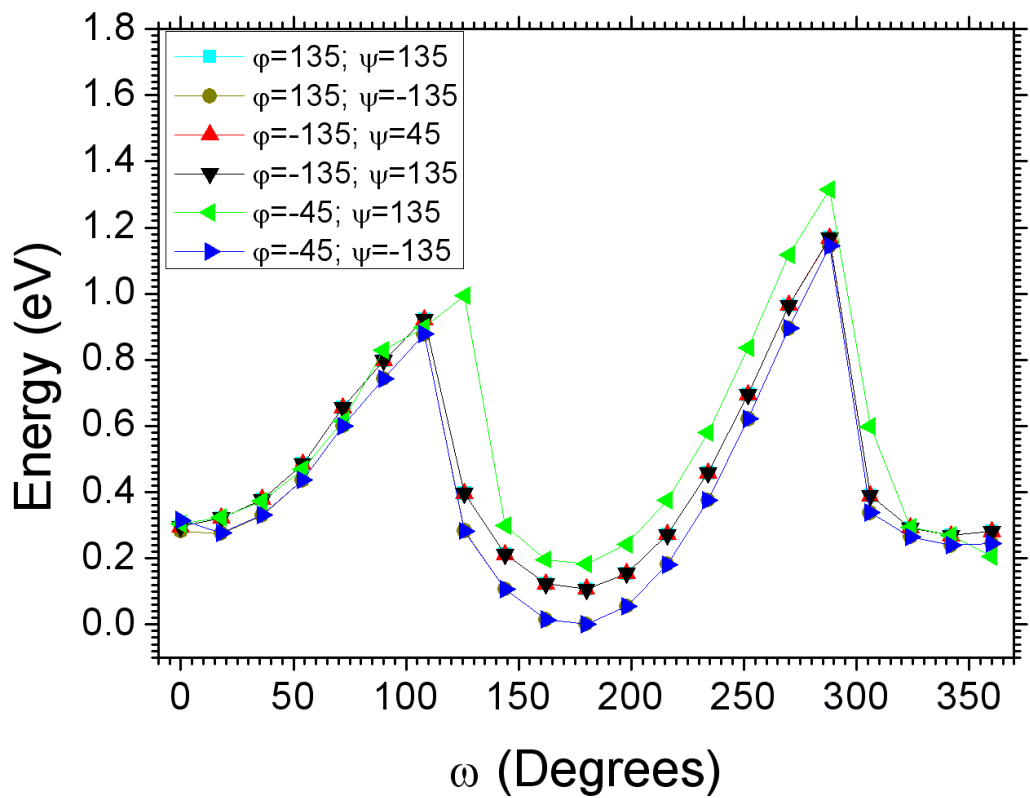}
\caption{Dependance of alanine tripeptides energy on angle $\omega$
calculated by the B3LYP/6-31G(d) method at different values of
angles $\varphi$ and $\psi$.} \label{omega_scan}
\end{figure}

For each amino acid there are only three dihedral angles formed by
atoms of the polypeptide chain which describe its twisting. The
angle $\omega$ (see fig. \ref{angles_def}) differs from the angles
$\varphi$ and $\psi$, because $C_i^{'}$ atom has the $sp^2$
hybridization state, what leads to formation of a quasi-double bond
between $C_i^{'}$ and $N_{i+1}$ atoms. Therefore, the angle $\omega$
is often referred as a "stiff" degree of freedom, whose value
depends only slightly on both the polypeptide constituent amino
acids and the values of other degrees of freedom. To illustrate this
fact, in figure \ref{omega_scan} we present the energy dependencies
on $\omega$ calculated for alanine dipeptide with different values
of angles $\varphi$ and $\psi$ in the cenral amino acid.

From this figure it is clear that there are two stable states in the
system with $\omega=0^{o}$ and $\omega=180^{o}$ which do not depend
on the angles $\varphi$ and $\psi$. The heights of the barriers
between these states are weakly depend on $\varphi$ and $\psi$,
being equal to $\sim$1 eV=23.06 kcal/mol.

The calculation shows that at temperatures close to the room
temperature, the value of the angle $\omega$ changes
insignificantly. The potential energy surface  as a function of the
angles $\varphi$ and $\psi$ appears to be much more complex as it is
shown in the next section.

\subsection{Potential energy surface for alanine tripeptide}

\begin{figure}[h]
\includegraphics[scale=0.8,clip]{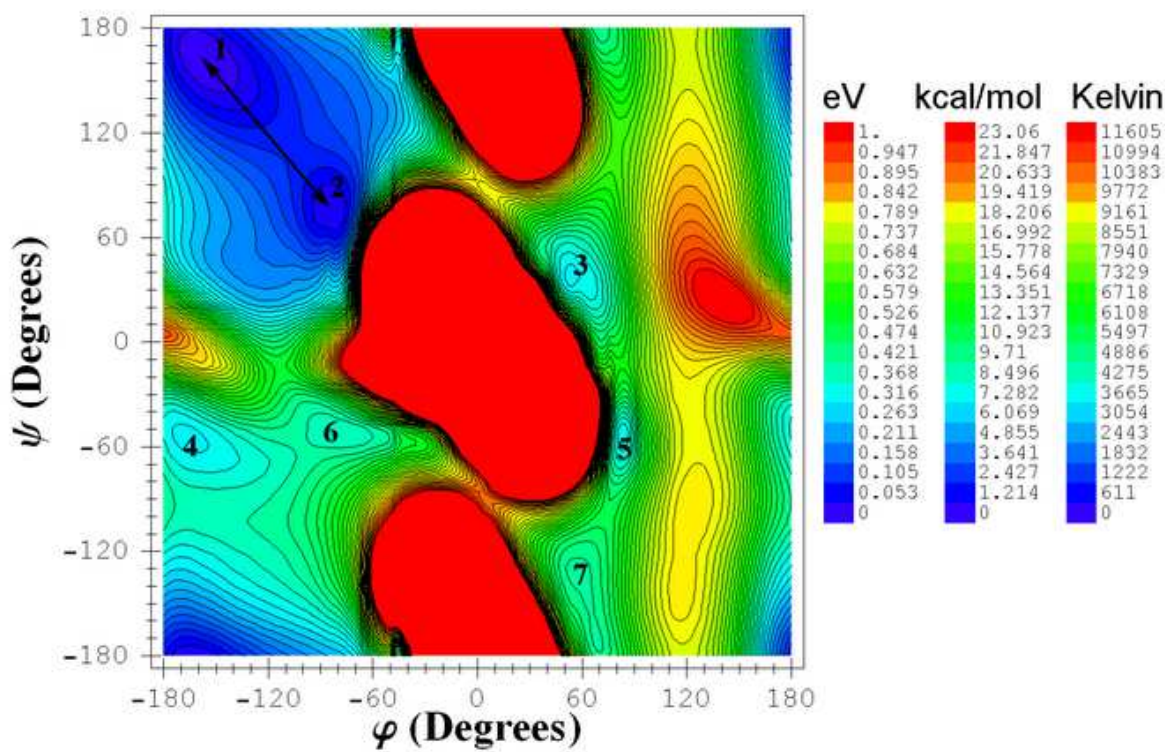}
\caption{Potential energy surface for the alanine tripeptide
calculated by the B3LYP/6-31G(2d,p) method. Energies are given in
eV, kcal/mol and Kelvin. Numbers mark energy minima on the potential
energy surface. Arrows show transition paths between different
conformations of the molecule.} \label{map_ala3}
\end{figure}

\begin{figure}[p]
\includegraphics[scale=0.79,clip]{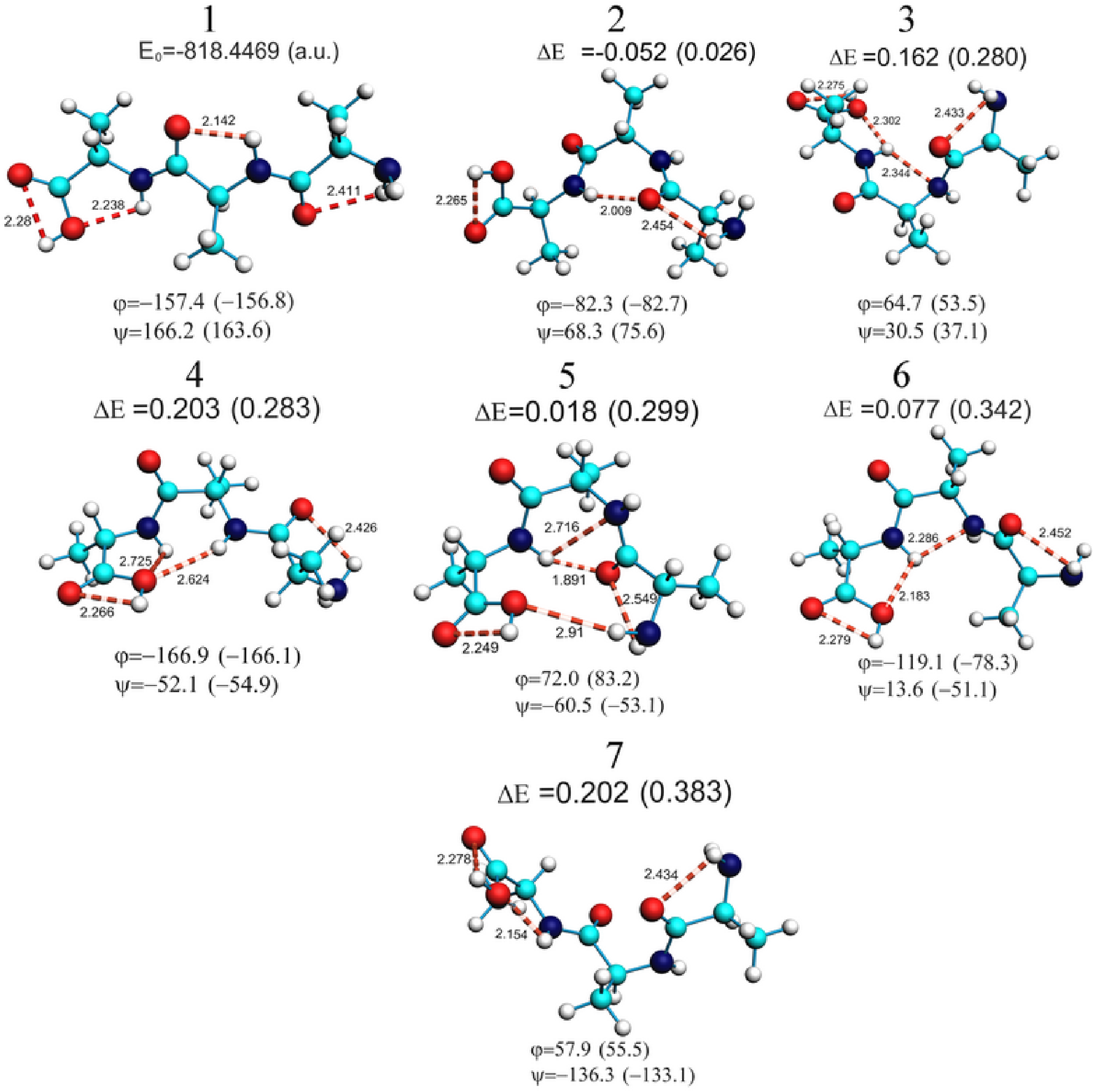}
\caption{Optimized conformations of the alanine tripeptide.
Different geometries correspond to different minima on the potential
energy surface (see contour plot in figure \ref{map_ala3}). Below
each image we present angles $\varphi$ and $\psi$, which have been
obtained with accounting for relaxation of all degrees of freedom in
the system. Values in brackets give the angles calculated without
accounting for relaxation. Above each image the energy of the
corresponding conformation is given in eV. The energies are counted
from the energy of conformation 1 (the energy of conformation 1 is
given in a.u.). Values in parentheses correspond to the energies
obtained without relaxation of all degrees of freedom in the system.
Dashed lines show the strongest hydrogen bonds. Their lengths are
given in angstroms.} \label{geom_ala3}
\end{figure}

In figure \ref{map_ala3} we present the potential energy surface for
the alanine tripeptide calculated by the B3LYP/6-31G(2d,p) method.
The energy scale is given in eV, kcal/mol and Kelvin. Energies on
the plot are measured from the lowest energy minimum of the
potential energy surface.

From the figure follows that there are several minima on the
potential energy surface. They are numbered according to the value
of the corresponding energy value. Each minimum corresponds to a
certain conformation of the molecule. These conformations differ
significantly from each other. In the case of alanine tripeptide
there are six conformations, shown in figure \ref{geom_ala3}. Dashed
lines show the strongest hydrogen bonds in the system, which arise
when the distance between hydrogen and oxygen atoms becomes less
then 2.9 angstroms.

To calculate the potential energy surface the following procedure
was adopted. Once the stable structure of the molecule has been
determined and optimized, all but two (these are the angles
$\varphi$ and $\psi$ in the central amino acid) degrees of freedom
were frozen. Then the energy of the molecule was calculated by
varying $\varphi$ and $\psi$. This procedure was used to calculate
all potential energy surfaces presented below in this section. It
allows one to find efficiently the minima on the energy surface and
to determine the main stable conformations of the molecule. The
absolute energy values of different conformations of the tripeptide
found by this method are not too accurate, because the method does
not account for the relaxation of other degrees of freedom in the
system. To calculate the potential energy surface with accounting
for the relaxation one needs 20-30 times more of the computer time.
Therefore, a calculations with accounting for the relaxation have
not been performed in our work. Instead, we have performed a
complete optimization of the molecular conformations, corresponding
to all minima on the calculated potential energy surface.

In figure \ref{geom_ala3} we compare stable conformations of the
alanine tripeptide calculated with and without accounting for the
relaxation of all atoms in the system. As it is seen from this
figure the angles $\varphi$ and $\psi$ differ by about 10 percent in
the two cases. This difference arises due to the coupling of
$\varphi$ and $\psi$ with other degrees of freedom. Note the change
of the sign of the relative energies of some conformations. This
effect is due to the rearrangement of side atoms (radicals) in the
polypeptide chain which lowers the energies of different
conformations differently.

In our work the potential energy surface has been calculated and
interpolated on the grid with the step of $18^{\circ}$. This step
size is an optimal one, because the interpolation error is about
$9^{\circ}$, i.e. comparable with the angle deviations caused by the
relaxation of all degrees of freedom in the system.

Note that for the alanine tripeptide an additional maximum appears
at $\varphi=120^{o}\pm 50^{o}$, $\psi=30^{o}\pm 30^{o}$, while it is
absent on the potential energy surface for the glycine tripeptide
\cite{twisting_preprint}. This maximum is a result of overlapping of
the side CH$_3$- radicals, which are substituted in the case of the
glycine polypeptide with the $H$- atoms.

\begingroup
\begin{table*}[h]
\caption{Comparison of dihedral angles $\varphi$ and $\psi$
corresponding to different conformations of alanine tripeptide.}
\label{tab:Ala3}

\begin{ruledtabular}
\begin{tabular}{c|cc|cc|cc}

     \multicolumn{1}{c|}{ Conformation } &
     \multicolumn{1}{c}{$\varphi$\cite{Head-Gordon91}} &
     \multicolumn{1}{c|}{$\psi$\cite{Head-Gordon91}} &
     \multicolumn{1}{c}{$\varphi$\cite{Gould94}} &
     \multicolumn{1}{c|}{$\psi$\cite{Gould94}}&
     \multicolumn{1}{c}{$\varphi$}&
     \multicolumn{1}{c}{$\psi$}\\

\hline
1 & -168.4 & 170.5 & -157.2  & 159.8 & -157.4  & 166.2  \\
2 & -      & -     & -60.7   & -40.7 & -82.3   & -68.3  \\
3 & 63.8   & 32.7  &  67.0   & 30.2  & 64.7    & 30.5   \\
4 & -      & -     &  -      & -     & -166.9  & -52.1  \\
5 & 74.1   & -57.3 &  76.0   & -55.4 &  72.0   & -60.5  \\
6 & -128.0 & 29.7  &  -130.9 & 22.3  &  -119.1 & 13.6   \\
7 & -      & -     &  -      & -     &  57.9   & -136.3 \\

\end{tabular}
\end{ruledtabular}
\end{table*}
\endgroup

In ref. \cite{Head-Gordon91} and ref. \cite{Gould94} several stable
conformations were found for alanine and glycine dipeptides. The
values of angles $\varphi$ and $\psi$ for the stable conformations
of dipeptide and tripeptide are close indicating that the third
amino acid in tripeptide makes relatively small influence on the
values of dihedral angles of two other amino acids. In earlier
papers refs. \cite{Head-Gordon91,Gould94} dipeptides were studied
within the framework of the Hartree-Fock theory. In ref.
\cite{Head-Gordon91}, values of $\varphi$ and $\psi$ were obtained
by the HF/6-31+G* method, and in ref. \cite{Gould94} by HF/6-31G**.
In table \ref{tab:Ala3} we compare the results of our calculation
for tripeptide with the corresponding data obtained for dipeptides.
Some discrepancy between the values presented is due to the
difference between the dipeptide and tripeptide (i.e. the third
alanine in the tripeptide affects the values of angles $\varphi$ and
$\psi$). However, another source of discrepancy might arise due to
accounting for the many-electron correlations in the DFT and
neglecting this effect in the Hartree-Fock theory used in refs.
\cite{Head-Gordon91,Gould94}.

\begin{figure}[h]
\includegraphics[scale=0.9,clip]{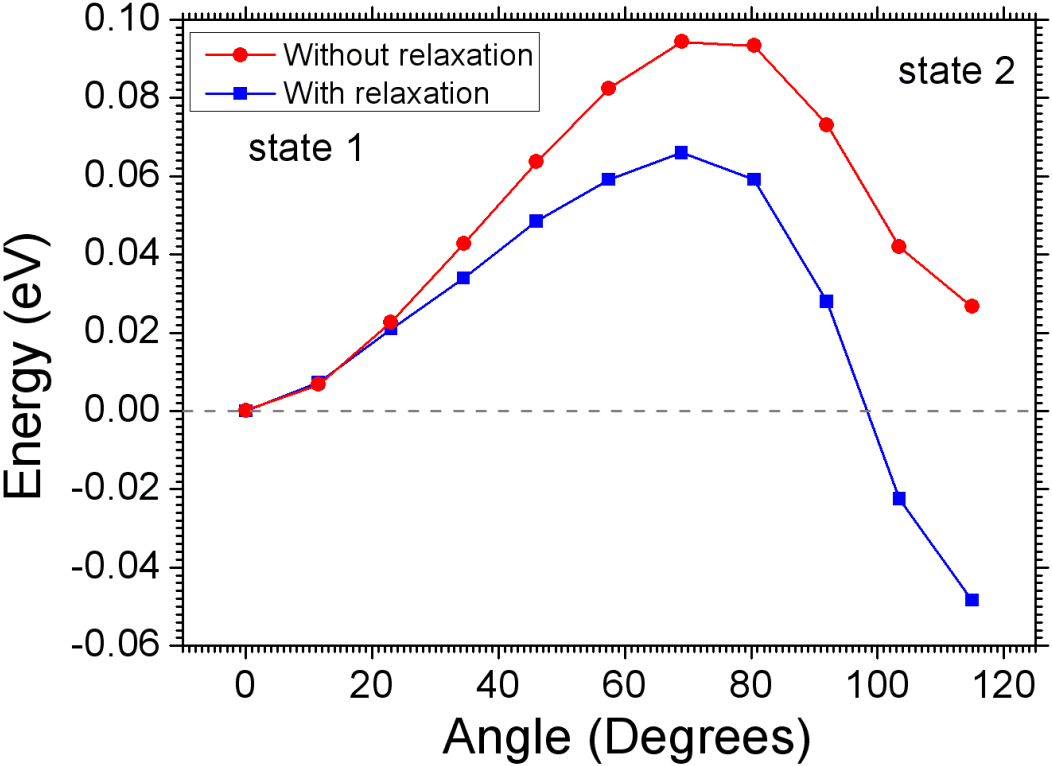}
\caption{Transition barriers for between conformations $1
\leftrightarrow 2$ of the alanine tripeptide. Circles and squares
correspond to the barriers calculated without and with relaxation of
all degrees of freedom in the system.} \label{barrier_ala3}
\end{figure}

Figure \ref{map_ala3} shows that some domains of the potential
energy surface, where the potential energy of the molecule increases
significantly, appear to be unfavorable for the formation of a
stable molecular configuration. The growth of energy takes place
when some atoms in the polypeptide chain approach each other at
small distances. Accounting for the molecule relaxation results in
the decrease of the system energy in such cases, but the resulting
molecular configurations remain unstable. We call such domains on
the potential energy surface as forbidden ones. In figure
\ref{map_ala3} one can identify two forbidden regions in the
vicinity of the points (0, 0) and (0, 180). At (0, 0)a pair of
hydrogen and oxygen atoms approach to the distances much smaller
than the characteristic $H-O$ bond length. This leads to a strong
interatomic repulsion caused by the exchange interaction of
electrons. At (0, 180) the Coulomb repulsion of pair of oxygen atoms
causes the similar effect.

Figure \ref{map_ala3} shows that there are six minima on the
potential energy surface for alanine tripeptide. The transition
barrier between the conformations $1 \leftrightarrow 2$ is shown in
figure \ref{barrier_ala3}.  The barrier has been calculated with and
without relaxation of the atoms in the system. The corresponding
transition path is marked in figure \ref{map_ala3} by an arrow. This
comparison demonstrates that accounting for the relaxation
significantly lowers the barrier height and influences the relative
value of energy of the minima.

Let us now estimate the time needed for a system for the transition
from one conformation to another. To do this we use the Arhenius
equation, which reads as:

\begin{equation}
\frac{1}{\tau}=\Omega e^{-\frac{\Delta E}{kT}} \label{Arhenius_eq}
\end{equation}

\noindent where $\tau$ is the transition time, $\Omega$ is the
factor, determining how frequently the system approaches the
barrier, $\Delta E$ is the barrier height, $T$  is the temperature
of the system, $k$  is the Bolzmann factor.

\begin{figure}[h]
\includegraphics[scale=0.8,clip]{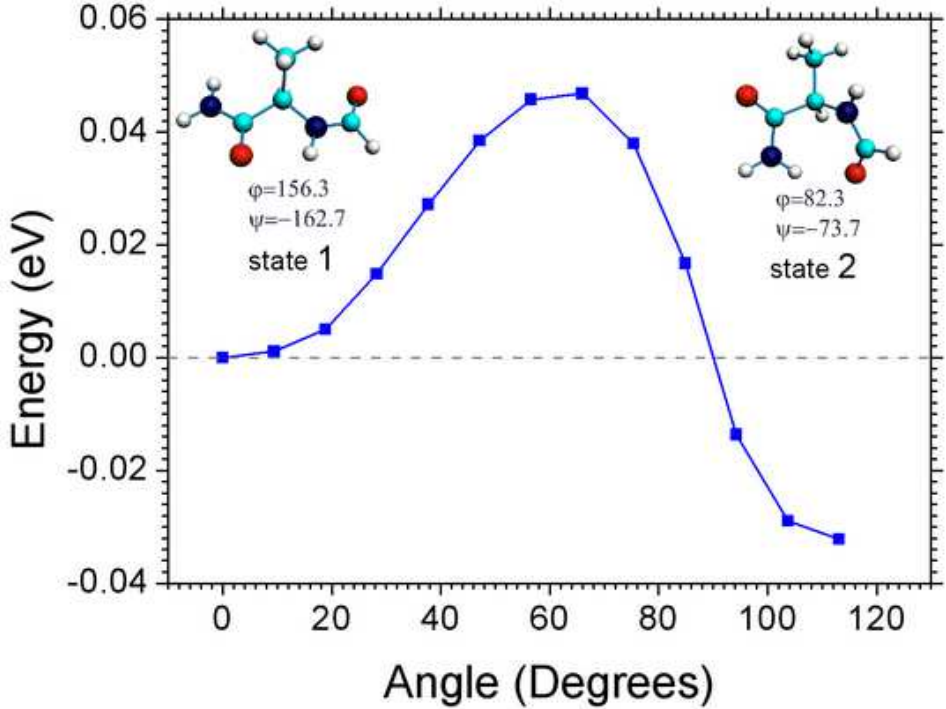}
\caption{Transition barriers for between conformations $1
\leftrightarrow 2$ of alanine dipeptide analog calculated by the
B3LYP/6-31+G(2d,p) method accounting for the relaxation of all
degrees of freedom in the system. Structure of the conformations 1
and 2 is shown near each minimum.} \label{barrier_gly2}
\end{figure}

Figure \ref{barrier_gly2} shows the transition barrier between two
main conformations of the alanine dipeptide analog
((S)-$\alpha$-(formylamino)propanamide). It is seen that $\Delta
E_{1\rightarrow2}=0.047$ eV for the transition $1\rightarrow2$,
while $\Delta E_{2\rightarrow1}=0.079$ eV for the transition
$2\rightarrow1$. The frequency $\Omega$ for this molecule is equal
to 42.87 cm$^{-1}$. Thus, at $T=300$ K, we obtain $\tau_{2\times
Ala}^{1\rightarrow2} \sim 5$ ps and $\tau_{2\times
Ala}^{2\rightarrow1} \sim 17$ ps. This result is in excellent
agreement with the molecular dynamics simulations results obtained
in ref. \cite{Salahub01} predicting $\tau\sim7$ ps for the
transition ${1\rightarrow2}$ and $\tau\sim19$ ps for the transition
${2\rightarrow1}$. This comparison demonstrates that our method is
reliable enough and it can be used for the estimation of transition
times between various conformations of the polypeptides.

Using the B3LYP/6-31G(2d,p) method we have calculated the
frequencies of normal vibration modes for the alanine tripeptide.
The characteristic frequency corresponding to twisting of the
polypeptide chain is equal to 32.04 cm$^{-1}$. From figure
\ref{barrier_ala3} follows that $\Delta E_{1\rightarrow2}=0.066$ eV
for the transition $1\rightarrow2$ and $\Delta
E_{2\rightarrow1}=0.114$ eV for the transition $2\rightarrow1$.
Thus, we obtain $\tau_{3\times Ala}^{1\rightarrow2} \sim 13$ ps and
$\tau_{3\times Ala}^{2\rightarrow1} \sim 86$ ps. Let us note that
these transition times can be measured experimentally by means of
NMR refs. \cite{Rubin04,Bax03}.

\subsection{Potential energy surface for alanine hexapeptide with
the sheet and the helix secondary structure}

\begin{figure}[b]
\includegraphics[scale=0.825,clip]{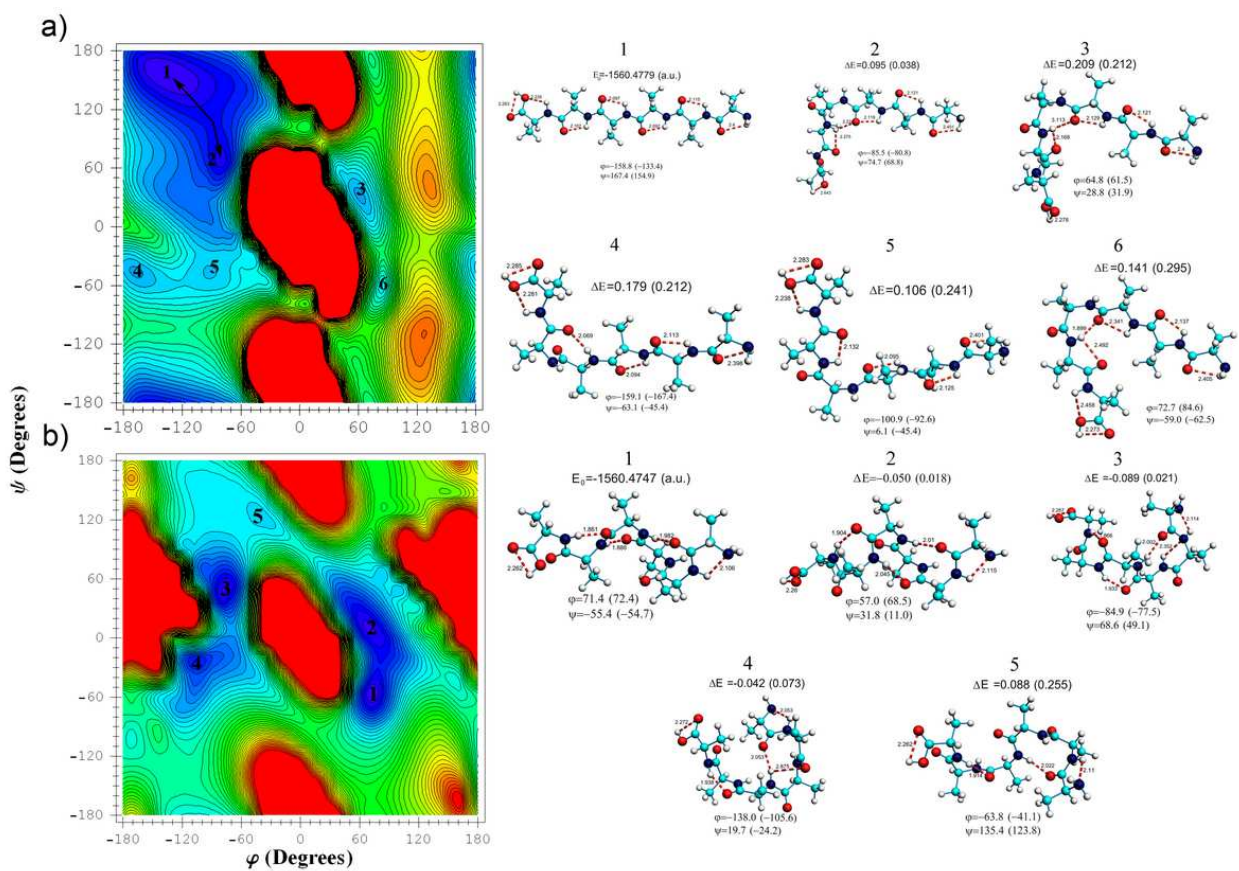}
\caption{Potential energy surface for the alanine hexapeptide with
the sheet secondary structure (part a) and with the helix secondary
structure (part b) calculated by the B3LYP/6-31G(2d,p) method.
Energy scale is given in figure \ref{map_ala3}. Numbers mark energy
minima on the potential energy surface. Images of optimized
conformations of the alanine hexapeptide are shown near the
corresponding energy landscape. Values of angles $\varphi$ and
$\psi$, as well as the relative energies of the conformations are
given analogously to that in figure \ref{geom_ala3}.} \label{ala6}
\end{figure}

\begin{figure}[h]
\includegraphics[scale=0.8,clip]{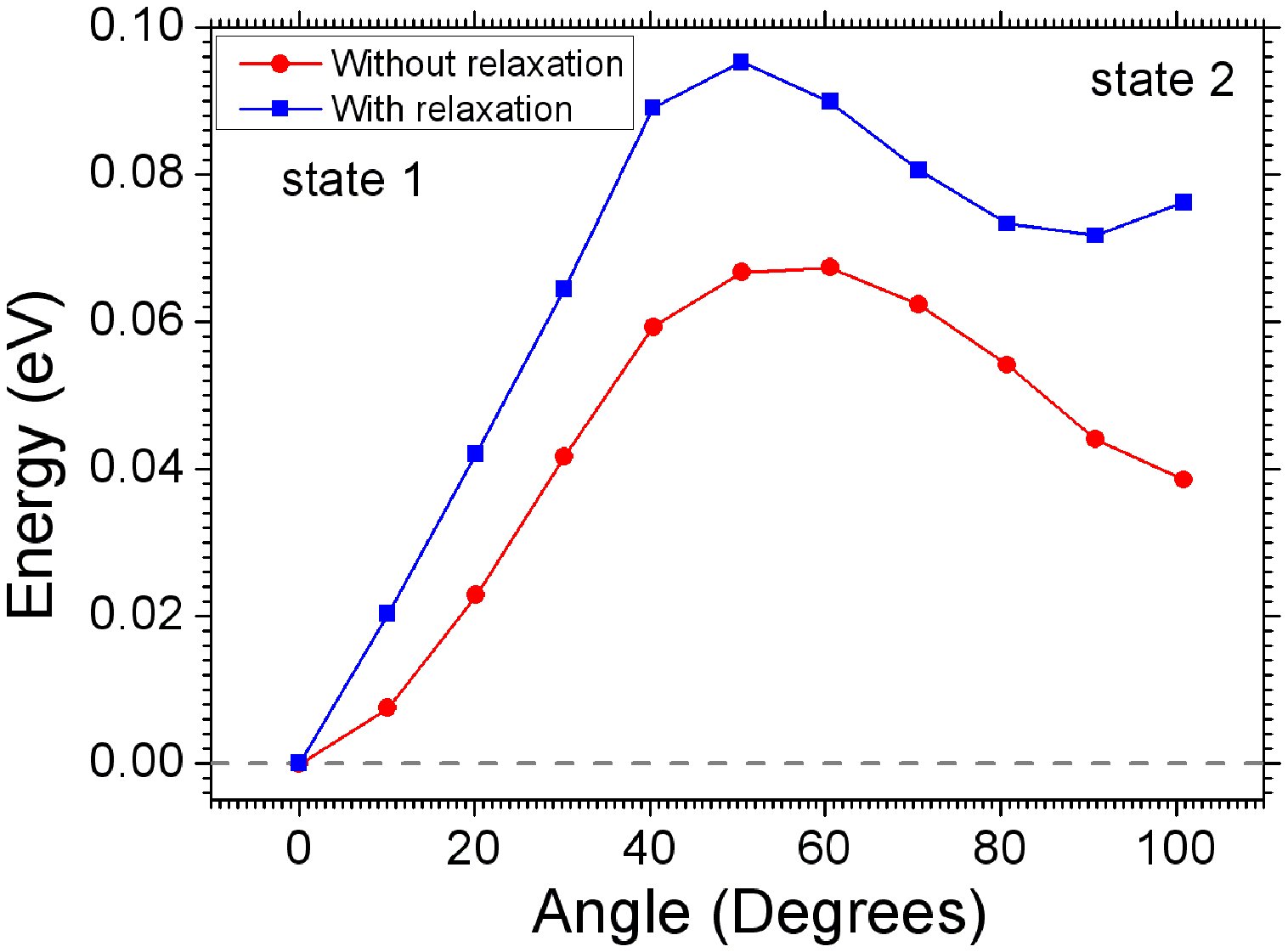}
\caption{Transitions barriers between conformations $1
\leftrightarrow 2$ of the alanine hexapeptide with the sheet
secondary structure. Circles and squares correspond to the barriers
calculated without and with relaxation of all degrees of freedom in
the system.} \label{barrier_ala6_sheet}
\end{figure}

In figure \ref{ala6} we present contour plots of the potential
energy surface for the alanine hexapeptide with the sheet (part a)
and the helix (part b) secondary structure respectively versus
dihedral angles $\varphi$ and $\psi$. In both cases the forbidden
regions arise because of the repulsion of oxygen and hydrogen atoms
analogously to the alanine tripeptide case.

Minima 1-6 on the potential energy surface \ref{ala6}a correspond to
different conformations of the alanine hexapeptide with the sheet
secondary structure. Note that these minima are also present on the
potential energy surface of the alanine tripeptide (see fig.
\ref{map_ala3}). Geometries of the conformations 1-6 are shown on
the right-hand side of figure \ref{ala6}a.

Energy barrier as a function of a scan variable (see figure
\ref{ala6}a) for the transition between conformations 1 and 2 is
shown in figure \ref{barrier_ala6_sheet}. The energy dependence has
been calculated with and without relaxation of all the atoms in the
system. In the case of alanine hexapeptide with the sheet secondary
structure the barrier height for the transition $1\rightarrow2$ is
significantly higher than for the transition $2\rightarrow1$, being
equal to 0.095 eV and 0.023 eV, respectively. The normal vibration
mode frequency, corresponding to the twisting of the polypeptide
chain is equal to 6.24 cm$^{-1}$ and was calculated with the
B3LYP/STO-3G method. Using equation (\ref{Arhenius_eq}) one derives
the transition times at room temperature: $\tau_{6\times
Gly}^{1\rightarrow2} \sim 211$ ps, $\tau_{6\times
Gly}^{2\rightarrow1} \sim 13\ $ps.

Let us now consider alanine hexapeptide with the helix secondary
structure. The potential energy surface for this polypeptide is
shown in figure \ref{ala6}b. The positions of minima on this surface
are shifted significantly compared to the cases discussed above.
This change takes place because of the influence of the secondary
structure of the polypeptide on the potential energy surface. The
geometries of the most stable conformations are shown on the
right-hand side of figure \ref{ala6}b.

For the alanine hexapeptide with the helix secondary structure there
is a maximum at $\varphi\sim180^{o}$ and $\psi\sim40^{o}$ in
addition to the central maxima on the potential energy surface. This
maximum appears because of the repulsive interaction of the
outermost amino acids side radicals.

It is worth noting that for some conformations of alanine
hexapeptide the angles $ \varphi $ and $ \psi $ change significantly
when the relaxation of all degrees of freedom in the system
accounting for (see for example conformations 1, 5 in fig. \ref
{ala6}a and conformations 2, 4 in fig. \ref {ala6}b). This means
that the potential energy surface of the alanine hexapeptide in the
vicinity of the mentioned minima is very sensitive to the relaxation
of all degrees of freedom. However, calculation of the potential
energy surface with accounting for the relaxation of all degrees of
freedom is unfeasible task. Indeed, one needs about 2000 hours of
computer time (Pentium Xeon 2.4 GHz) for the calculation of the
potential energy surface for the alanine hexapeptide. To perform an
analogues calculation with accounting for the relaxation about 5
years of computer time would be needed. Nevertheless, the potential
energy surface calculated without accounting for the relaxation
carries a lot of useful information. Thus, one can predetermine
stable conformations of polypeptide, which then can be used as
starting configurations for further energy minimization.

\subsection{Comparison of calculation results with experimental data}

Nowadays, the structure of many proteins has  been determined
experimentally ref. \cite{Protbase}. Knowing the protein structure
one can find the angles $ \varphi $ and $ \psi $ for each amino acid
in the protein.

\begin{figure}[ht]
\includegraphics[scale=0.75,clip]{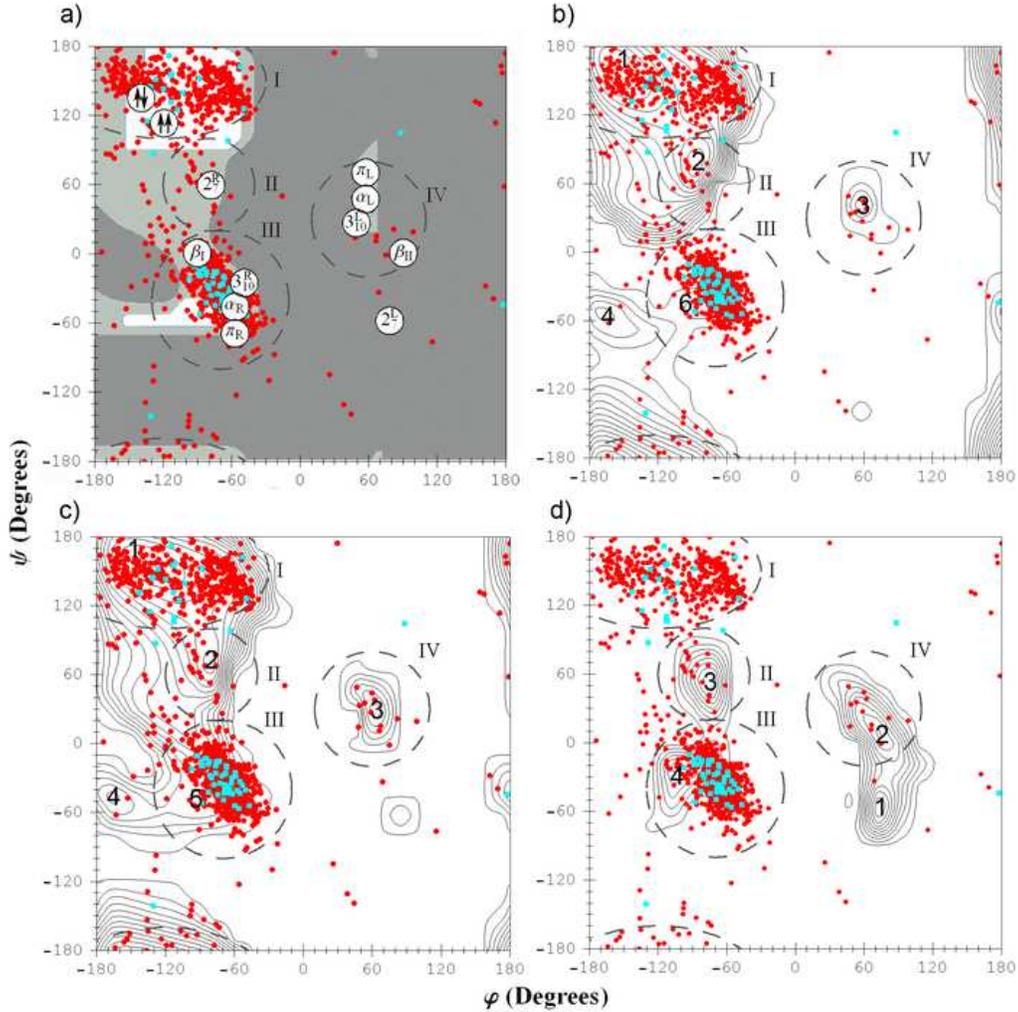}
\caption{Comparison of angles $\varphi$ and $\psi$ of alanine
residues in protein structures selected from the Brookhaven Protein
Data Bank \cite{Protbase,RamachandranPlot} with the steric diagram
for poly-alanine \cite{Biochemistry_book} (part a)). Comparison of
angles $\varphi$ and $\psi$ of alanine residues in protein
structures selected from the Brookhaven Protein Data Bank
\cite{Protbase,RamachandranPlot} with the minima on the calculated
potential energy surfaces for: alanine tripeptide (b); alanine
hexapeptide in sheet conformation (c); alanine hexapeptide in helix
conformation (d). Transparent rhomboids correspond to alanines
surrounded with alanines, while filled circles correspond to
alanines surrounded by other amino acids. Dashed ellipses mark the
regions of higher concentration of the observed angles.}
\label{Ala_experiment}
\end{figure}

In figure \ref{Ala_experiment}a we show a map of the allowed and
forbidden conformations for alanine residues in poly-alanine chain
taken from ref. \cite{Biochemistry_book} (steric Ramachandran
diagram). This map was obtained from pure geometrical
considerations, in which the structure of the polypeptide was
assumed to be fixed and defined by the interatomic van der Waals
interaction radii. Depending on the distances between the atoms one
could distinguish three regions: completely allowed, conventionally
allowed and forbidden. The conformation is called completely allowed
if all the distances between atoms of different amino acids are
larger than some critical value $r_{ij}\ge r_{max}$. Conventionally
allowed regions on the potential energy surface correspond to the
conformations of the polypeptide, in which the distances between
some atoms of different amino acids lie within the interval
$r_{min}\le r_{ij} < r_{max}$. All other conformations are referred
to as forbidden. The values of $r_{min}$ and $r_{max}$ are defined
by the types of interacting atoms and can be found in the textbooks
(see, e.g., \cite{Biochemistry_book}). In figure
\ref{Ala_experiment}a we mark the completely allowed regions with
white, the conventionally allowed regions with light gray and the
forbidden regions with dark gray color. In this figure we mark the
points, which correspond to the geometries of alanine, whose
periodical iteration leads to the formation of chains with specific
secondary structure. In table \ref{tab:RamAngl} we compile the
values of angles $\varphi$ and $\psi$, which correspond to the most
prominent poly-alanine secondary structures. For the illustrative
purposes we mark these points by white circles with the
corresponding type of the secondary structure typed in. Thus,
$2_7^{R}$, $2_7^{L}$ are the right-handed and the left-handed $2_7$
helix; $3_{10}^{R}$, $3_{10}^{L}$ are the right-handed and the
left-handed $3_{10}$ helix; $\alpha_{R}$, $\alpha_{L}$ are the
right-handed and the left-handed $\alpha-$helix ($4_{13}$);
$\pi_{R}$, $\pi_{L}$ are the right-handed and the left-handed
$\pi-$helix ($5_{16}$); $\uparrow\uparrow$, $\uparrow\downarrow$ are
the parallel and antiparallel $\beta$ sheets. $\beta_{I}$,
$\beta_{II}$ correspond to the $\beta-$turns of types I and II
respectively.

\begingroup
\begin{table*}[h]
\caption{Angles $\varphi$ and $\psi$ corresponding to the most
prominent poly-alanine secondary structures.} \label{tab:RamAngl}

\begin{ruledtabular}
\begin{tabular}{ccc}

     \multicolumn{1}{c}{Structure type} &
     \multicolumn{1}{c}{$\varphi$ (Deg.)} &
     \multicolumn{1}{c}{$\psi$ (Deg.)}\\

\hline
right-handed (left-handed) $2_7$ helix    & -78 (78) & 59 (-59)  \\
right-handed (left-handed) $3_{10}$ helix & -49 (49) & -26 (26)  \\
right-handed (left-handed) $\alpha-$helix ($4_{13}$) & -57 (57) & -47 (47)  \\
right-handed (left-handed) $\pi-$helix ($5_{16}$)    & -57 (57) & -70 (70)  \\
parallel $\beta$ sheet ($\uparrow\uparrow$) & -119 & 113 \\
antiparallel $\beta$ sheet ($\uparrow\downarrow$) & -139 & 135 \\
$\beta-$turn of type I & -90 & 0 \\
$\beta-$turn of type II & 90 & 0 \\

\end{tabular}
\end{ruledtabular}
\end{table*}
\endgroup

Note that not all of the structures listed above are present equally
in proteins. In figure \ref{Ala_experiment}a we show the
distribution of the angles $\varphi$ and $\psi$ of alanine residues
in protein structures selected from the Brookhaven Protein Data Bank
\cite{Protbase,RamachandranPlot}. It is possible to distinguish four
main regions, in which most of experimental points are located. In
figure \ref{Ala_experiment} these regions are schematically shown
with dashed ellipses. Note, that these ellipses are used for
illustrative purposes only, and serve for a better understanding of
the experimental data. The regions in which most of the observed
angles $\varphi$ and $\psi$ are located correspond to different
secondary structures of the poly-alanine. Thus, region I corresponds
to the parallel and antiparallel $\beta-$sheets. Region II
corresponds to the right-handed $2_7^R$ helix.  Region III
corresponds to the right-handed $\alpha_R-$helix, right-handed
$\pi_R-$helix, right-handed $3_{10}^R$ helix and $\beta-$turn of
type I. Region IV corresponds to the left-handed $\alpha_L-$helix,
right-handed $\pi_L-$helix, left-handed $3_{10}^L$ helix and
$\beta-$turn of type II. In some cases there are several types of
secondary structure within one domain. In the present work we have
not studied the secondary structure of proteins systematically
enough to establish the univocal correspondence of the observed
experimental points to different types of the secondary structure.

Let us now compare the distribution of angles $\varphi$ and $\psi$
experimentally observed for proteins with the potential energy
landscape calculated for alanine polypeptides and establish
correspondence of the secondary structure of the calculated
conformations with the predictions of the simple Ramachandran model.

Region I corresponds to the minimum 1 on the both potential energy
surfaces of the alanine tripeptide (fig. \ref{Ala_experiment}b) and
the alanine hexpeptide with the secondary structure of sheet (fig.
\ref{Ala_experiment}c). These conformations correspond exactly to
the alanine chains in the $\beta$-sheet conformation (see fig.
\ref{geom_ala3} and \ref{ala6}a). Note that there is no minimum in
that region of the potential energy surface for alanine hexapeptide
with the secondary structure of helix (see fig.
\ref{Ala_experiment}d).

Region II corresponds to the minimum 2 on the both potential energy
surfaces \ref{Ala_experiment}b and \ref{Ala_experiment}c, as well as
to the minimum 3 on the potential energy surface
\ref{Ala_experiment}d. On the steric diagram for poly-alanine this
region corresponds to the right-handed $2_7^R$ helix. The structure
of conformations 2 on the surfaces \ref{Ala_experiment}b and
\ref{Ala_experiment}c differs from the structure of this particular
helix type. Only the central alanines, for which the angles
$\varphi$ and $\psi$ in figures \ref{Ala_experiment}b and
\ref{Ala_experiment}c are defined, have the structure of $2_7^R$
helix. Thus, one can refer to the conformations 2 as to the mixed
states, where the central part of the polypeptide chain has the
conformation of helix and the outermost parts have the conformation
of sheet. Conformation 3 on the surface \ref{Ala_experiment}d is
also a mixed state. Here one can distinguish one turn of $3_{10}^R$
helix and two turns of $2_7^R$ helix (see fig. \ref{ala6}b).

Region III corresponds to the structure of right-handed
$\alpha_R-$helix, right-handed $3_{10}^R$ helix, right-handed
$\pi_R-$helix and $\beta-$turn. It corresponds to minima 6, 5 and 4
on the potential energy surfaces \ref{Ala_experiment}b,
\ref{Ala_experiment}c and \ref{Ala_experiment}d respectively.
Conformation 6 can not be assigned to any specific type of secondary
structure because the chain is too short. Note, that conformation 6
is even not a stable one on the potential energy surface of the
alanine tripeptide. The most probable types of secondary structures
in that region of the potential energy surface are right-handed
$\alpha_R-$helix and $\beta-$turn. However, for the formation of a
single turn of  $\alpha_R-$helix (or for the formation of
$\beta-$turn) at least four amino acids are needed. Conformation 5
on the potential energy surface of the alanine hexapeptide can be
characterized as a partially formed $\beta-$turn because the
alanine, for which the dihedral angles $\varphi$ and $\psi$ in
figure \ref{Ala_experiment}c are defined has the geometry of
$\beta-$turn, but its neighbor forms a $\beta-$sheet (see fig.
\ref{ala6}a). Conformation 4 on the potential energy surface
\ref{ala6}b changes significantly after accounting for the
relaxation of all degrees of freedom in the system, and gets outside
the region III. In this conformation one can locate fragments of
right-handed $2_7^R$ and $3_{10}^R$ helixes. The point corresponding
to the minimum 4 (after accounting for the relaxation) lies outside
regions II and III because angles $\varphi$ and $\psi$ in figure
\ref{Ala_experiment}d are defined for the amino acid between two
helix fragments.

Region IV is represented by the structure of left-handed
$\alpha_L-$helix, left-handed $3_{10}^L$ helix, left-handed
$\pi_L-$helix and $\beta-$turn of type II. The fragments with those
types of secondary structures are very rare met in native proteins.
To form these structures it is necessary to have at least four amino
acids, therefore minima 3 on the potential energy surface for
alanine tripeptide can not be compared to any type of the mentioned
secondary structures. Region IV corresponds to the conformations 3
and 2 on the surfaces \ref{Ala_experiment}c and
\ref{Ala_experiment}d respectively. Conformation 3 on the surface
\ref{Ala_experiment}c corresponds to partially formed $\beta-$turn,
because the alanine, for which the dihedral angles $\varphi$ and
$\psi$ in figure \ref{Ala_experiment}c are plotted has the
configuration of $\beta-$turn but the neighboring amino acid in the
polypeptide chain forms $\beta-$sheet (see fig. \ref{ala6}a).
Conformation 2 on the potential energy surface \ref{Ala_experiment}d
lies outside the region IV, but accounting for the relaxation of all
degrees of freedom shifts the minimum on the potential energy
surface to the allowed region of left-handed $\alpha_L-$ and
$3_{10}^L$ helix (see fig. \ref{ala6}b). The geometry of
conformation 2 is similar to the geometry of left-handed $3_{10}^L$
helix (see fig. \ref{ala6}b). The main differences in the structure
are caused by the insufficient length of the polypeptide chain to
form a regular helix structure.

%Finally we mention that the conformation 1 on the potential energy surface
%\ref{Ala_experiment}d corresponds to the structure of left-handed
%$2_7^L$ helix on the steric diagram \ref{Ala_experiment}a, being a
%mixed state (here one can distinguish two turns of left-handed
%$2_7^L$ helix and one turn of left-handed $3_{10}^L$ helix). This
%happens because the angles $\varphi$ and $\psi$ in figure
%\ref{Ala_experiment}d are defined for one of the alanines from
%$2_7^L$ helix fragment.

\section{Conclusion}
\label{conclusions}

In the present paper the multidimensional potential energy surfaces
for amino acid chains consisting of three and six alanines has been
investigated and the conformational properties of these systems with
respect to the twisting of the polypeptide chain have been
described. The calculations have been carried out within {\it ab
initio} theoretical framework based on the density functional theory
and accounting for all the electrons in the system. We have
determined stable conformations and calculated the energy barriers
for transitions between them. Using a thermodynamic approach, we
have estimated times of the characteristic transitions between the
conformations. It was demonstrated that the transition times lie
within the picosecond region. Our estimates are compared with the
available molecular-dynamics simulations results, and the
correspondence between the results of the two different methods is
reported. A strong barrier asymmetry between neighboring stable
conformations on the potential energy surface was found.

We compared for the first time values of angles $\varphi$ and $\psi$
for alanine residues experimentally observed in real proteins with
the coordinates of minima on the potential energy surfaces. This
comparison showed that all profound minima on the potential energy
surfaces correspond to the regions in which experimentally measured
values of $\varphi$ and $\psi$ are located. We have also analysed
how the secondary structure of polypeptide chains influences the
formation of the potential energy landscapes. For the chains of six
amino acids with the secondary structures of sheet and helix the
influence of the secondary structure on the stable conformations of
the molecule has been demonstrated.

The results of this work can be utilized for modeling more complex
molecular systems. For example, the suggested model for for the
estimation of the characteristic transition times can be used for
longer polypeptide chains, also consisting of different amino acids
and for estimates of time of proteins folding. It is also possible
to use the results of the present work for testing the applicability
and accuracy of different model approaches for the polypeptide
description requiring much less computer time than {\it ab initio}
calculations.

\section{Acknowledgements}
This work is partially supported by the European Commission within
the Network of Excellence project EXCELL, by INTAS under the grant
03-51-6170 and by the Russian Foundation for Basic Research under
the grant 06-02-17227-a. We are grateful to Dr. A. Korol and Dr. O.
Obolensky for their help in preparation of this manuscript. The
possibility to perform complex computer simulations at the Frankfurt
Center for Scientific Computing is also gratefully acknowledged.


\begin{thebibliography}{41}
\expandafter\ifx\csname natexlab\endcsname\relax\def\natexlab#1{#1}\fi
\expandafter\ifx\csname bibnamefont\endcsname\relax
  \def\bibnamefont#1{#1}\fi
\expandafter\ifx\csname bibfnamefont\endcsname\relax
  \def\bibfnamefont#1{#1}\fi
\expandafter\ifx\csname citenamefont\endcsname\relax
  \def\citenamefont#1{#1}\fi
\expandafter\ifx\csname url\endcsname\relax
  \def\url#1{\texttt{#1}}\fi
\expandafter\ifx\csname urlprefix\endcsname\relax\def\urlprefix{URL }\fi
\providecommand{\bibinfo}[2]{#2}
\providecommand{\eprint}[2][]{\url{#2}}

\bibitem[{\citenamefont{Karas and Hillenkamp}(1988)}]{Karas88}
\bibinfo{author}{\bibfnamefont{M.}~\bibnamefont{Karas}} \bibnamefont{and}
  \bibinfo{author}{\bibfnamefont{F.}~\bibnamefont{Hillenkamp}},
  \bibinfo{journal}{Anal. Chem.} \textbf{\bibinfo{volume}{60}},
  \bibinfo{pages}{2299} (\bibinfo{year}{1988}).

\bibitem[{\citenamefont{Hillenkamp and Karas}(2000)}]{Karas00}
\bibinfo{author}{\bibfnamefont{F.}~\bibnamefont{Hillenkamp}} \bibnamefont{and}
  \bibinfo{author}{\bibfnamefont{M.}~\bibnamefont{Karas}},
  \bibinfo{journal}{Int. J. of Mass Spect.} \textbf{\bibinfo{volume}{200}},
  \bibinfo{pages}{71} (\bibinfo{year}{2000}).

\bibitem[{\citenamefont{Karas et~al.}(2003)\citenamefont{Karas, Bahr, Fournier,
  Gluckmann, and Pfenninger}}]{Karas03}
\bibinfo{author}{\bibfnamefont{M.}~\bibnamefont{Karas}},
  \bibinfo{author}{\bibfnamefont{U.}~\bibnamefont{Bahr}},
  \bibinfo{author}{\bibfnamefont{I.}~\bibnamefont{Fournier}},
  \bibinfo{author}{\bibfnamefont{M.}~\bibnamefont{Gluckmann}},
  \bibnamefont{and}
  \bibinfo{author}{\bibfnamefont{A.}~\bibnamefont{Pfenninger}},
  \bibinfo{journal}{J. of Mass Spect.} \textbf{\bibinfo{volume}{226}},
  \bibinfo{pages}{239} (\bibinfo{year}{2003}).

\bibitem[{\citenamefont{Wind and Lehmann}(2004)}]{Wind04}
\bibinfo{author}{\bibfnamefont{M.}~\bibnamefont{Wind}} \bibnamefont{and}
  \bibinfo{author}{\bibfnamefont{W.}~\bibnamefont{Lehmann}},
  \bibinfo{journal}{J. Anal. At. Spect.} \textbf{\bibinfo{volume}{19}},
  \bibinfo{pages}{20} (\bibinfo{year}{2004}).

\bibitem[{\citenamefont{Fenn et~al.}(1989)\citenamefont{Fenn, Mann, Meng, Wong,
  and Whitehouse}}]{Fenn89}
\bibinfo{author}{\bibfnamefont{J.}~\bibnamefont{Fenn}},
  \bibinfo{author}{\bibfnamefont{M.}~\bibnamefont{Mann}},
  \bibinfo{author}{\bibfnamefont{C.}~\bibnamefont{Meng}},
  \bibinfo{author}{\bibfnamefont{S.}~\bibnamefont{Wong}}, \bibnamefont{and}
  \bibinfo{author}{\bibfnamefont{C.}~\bibnamefont{Whitehouse}},
  \bibinfo{journal}{Science} \textbf{\bibinfo{volume}{246}},
  \bibinfo{pages}{64} (\bibinfo{year}{1989}).

\bibitem[{\citenamefont{ndsted Nielsen et~al.}(2004)\citenamefont{ndsted
  Nielsen, Andersen, Hvelplund, Liu, and Tomita}}]{Hvelplund04}
\bibinfo{author}{\bibfnamefont{S.~B.} \bibnamefont{ndsted Nielsen}},
  \bibinfo{author}{\bibfnamefont{J.}~\bibnamefont{Andersen}},
  \bibinfo{author}{\bibfnamefont{P.}~\bibnamefont{Hvelplund}},
  \bibinfo{author}{\bibfnamefont{B.}~\bibnamefont{Liu}}, \bibnamefont{and}
  \bibinfo{author}{\bibfnamefont{S.}~\bibnamefont{Tomita}},
  \bibinfo{journal}{J. Phys. B: At. Mol. Opt. Phys.}
  \textbf{\bibinfo{volume}{37}}, \bibinfo{pages}{R25} (\bibinfo{year}{2004}).

\bibitem[{\citenamefont{Finkelstein and Ptizin}(2002)}]{Ptizin_book}
\bibinfo{author}{\bibfnamefont{A.}~\bibnamefont{Finkelstein}} \bibnamefont{and}
  \bibinfo{author}{\bibfnamefont{O.}~\bibnamefont{Ptizin}},
  \emph{\bibinfo{title}{Physics of Proteins}} (\bibinfo{publisher}{Moscow
  University Press "Universitet"}, \bibinfo{year}{2002}).

\bibitem[{\citenamefont{M\"ulberg}(2004)}]{Muelberg_book}
\bibinfo{author}{\bibfnamefont{A.}~\bibnamefont{M\"ulberg}},
  \emph{\bibinfo{title}{Protein Folding}} (\bibinfo{publisher}{St. Petersburg
  University Press}, \bibinfo{year}{2004}).

\bibitem[{\citenamefont{Berman et~al.}(2000)\citenamefont{Berman, Westbrook,
  Feng, Gilliland, Bhat, Weissig, Shindyalov, and Bourne}}]{Protbase}
\bibinfo{author}{\bibfnamefont{H.}~\bibnamefont{Berman}},
  \bibinfo{author}{\bibfnamefont{J.}~\bibnamefont{Westbrook}},
  \bibinfo{author}{\bibfnamefont{Z.}~\bibnamefont{Feng}},
  \bibinfo{author}{\bibfnamefont{G.}~\bibnamefont{Gilliland}},
  \bibinfo{author}{\bibfnamefont{T.}~\bibnamefont{Bhat}},
  \bibinfo{author}{\bibfnamefont{H.}~\bibnamefont{Weissig}},
  \bibinfo{author}{\bibfnamefont{I.}~\bibnamefont{Shindyalov}},
  \bibnamefont{and} \bibinfo{author}{\bibfnamefont{P.}~\bibnamefont{Bourne}},
  \bibinfo{journal}{Nucleic Acids Research} \textbf{\bibinfo{volume}{28}},
  \bibinfo{pages}{235} (\bibinfo{year}{2000}).

\bibitem[{\citenamefont{Rubin}(2004)}]{Rubin04}
\bibinfo{author}{\bibfnamefont{A.}~\bibnamefont{Rubin}},
  \emph{\bibinfo{title}{Biophysics: Theoretical Biophysics}}
  (\bibinfo{publisher}{Moscow University Press "Nauka"}, \bibinfo{year}{2004}).

\bibitem[{\citenamefont{Head-Gordon et~al.}(1991)\citenamefont{Head-Gordon,
  Head-Gordon, Frisch, III, and Pople}}]{Head-Gordon91}
\bibinfo{author}{\bibfnamefont{T.}~\bibnamefont{Head-Gordon}},
  \bibinfo{author}{\bibfnamefont{M.}~\bibnamefont{Head-Gordon}},
  \bibinfo{author}{\bibfnamefont{M.}~\bibnamefont{Frisch}},
  \bibinfo{author}{\bibfnamefont{C.~B.} \bibnamefont{III}}, \bibnamefont{and}
  \bibinfo{author}{\bibfnamefont{J.}~\bibnamefont{Pople}}, \bibinfo{journal}{J.
  Am. Chem. Soc.} \textbf{\bibinfo{volume}{113}}, \bibinfo{pages}{5989}
  (\bibinfo{year}{1991}).

\bibitem[{\citenamefont{Gould et~al.}(1994)\citenamefont{Gould, Cornell, and
  Hillier}}]{Gould94}
\bibinfo{author}{\bibfnamefont{I.}~\bibnamefont{Gould}},
  \bibinfo{author}{\bibfnamefont{W.}~\bibnamefont{Cornell}}, \bibnamefont{and}
  \bibinfo{author}{\bibfnamefont{I.}~\bibnamefont{Hillier}},
  \bibinfo{journal}{J. Am. Chem. Soc} \textbf{\bibinfo{volume}{116}},
  \bibinfo{pages}{9250} (\bibinfo{year}{1994}).

\bibitem[{\citenamefont{Wang and Duan}(2004)}]{Zhi-Xiang04}
\bibinfo{author}{\bibfnamefont{Z.}~\bibnamefont{Wang}} \bibnamefont{and}
  \bibinfo{author}{\bibfnamefont{Y.}~\bibnamefont{Duan}}, \bibinfo{journal}{J.
  Comp. Chem.} \textbf{\bibinfo{volume}{25}}, \bibinfo{pages}{1699}
  (\bibinfo{year}{2004}).

\bibitem[{\citenamefont{Percel et~al.}(2003)\citenamefont{Percel, Farkas,
  J\'akli, Topol, and Csizmadia}}]{Percel03}
\bibinfo{author}{\bibfnamefont{A.}~\bibnamefont{Percel}},
  \bibinfo{author}{\bibfnamefont{O.}~\bibnamefont{Farkas}},
  \bibinfo{author}{\bibfnamefont{I.}~\bibnamefont{J\'akli}},
  \bibinfo{author}{\bibfnamefont{I.}~\bibnamefont{Topol}}, \bibnamefont{and}
  \bibinfo{author}{\bibfnamefont{I.}~\bibnamefont{Csizmadia}},
  \bibinfo{journal}{J. Comp. Chem.} \textbf{\bibinfo{volume}{24}},
  \bibinfo{pages}{1026} (\bibinfo{year}{2003}).

\bibitem[{\citenamefont{H\'udaky et~al.}(2004)\citenamefont{H\'udaky, H\'udaky,
  and Percel}}]{Hudaky04}
\bibinfo{author}{\bibfnamefont{I.}~\bibnamefont{H\'udaky}},
  \bibinfo{author}{\bibfnamefont{P.}~\bibnamefont{H\'udaky}}, \bibnamefont{and}
  \bibinfo{author}{\bibfnamefont{A.}~\bibnamefont{Percel}},
  \bibinfo{journal}{J. Comp. Chem.} \textbf{\bibinfo{volume}{25}},
  \bibinfo{pages}{1522} (\bibinfo{year}{2004}).

\bibitem[{\citenamefont{Improta and Barone}(2004)}]{Improta04}
\bibinfo{author}{\bibfnamefont{R.}~\bibnamefont{Improta}} \bibnamefont{and}
  \bibinfo{author}{\bibfnamefont{V.}~\bibnamefont{Barone}},
  \bibinfo{journal}{J. Comp. Chem.} \textbf{\bibinfo{volume}{25}},
  \bibinfo{pages}{1333} (\bibinfo{year}{2004}).

\bibitem[{\citenamefont{Vargas et~al.}(2002)\citenamefont{Vargas, Garza, Hay,
  and Dixon}}]{Vargas02}
\bibinfo{author}{\bibfnamefont{R.}~\bibnamefont{Vargas}},
  \bibinfo{author}{\bibfnamefont{J.}~\bibnamefont{Garza}},
  \bibinfo{author}{\bibfnamefont{B.}~\bibnamefont{Hay}}, \bibnamefont{and}
  \bibinfo{author}{\bibfnamefont{D.}~\bibnamefont{Dixon}}, \bibinfo{journal}{J.
  Phys. Chem. A} \textbf{\bibinfo{volume}{106}}, \bibinfo{pages}{3213}
  (\bibinfo{year}{2002}).

\bibitem[{\citenamefont{Kashner and Hohl}(1998)}]{Kashner98}
\bibinfo{author}{\bibfnamefont{R.}~\bibnamefont{Kashner}} \bibnamefont{and}
  \bibinfo{author}{\bibfnamefont{D.}~\bibnamefont{Hohl}}, \bibinfo{journal}{J.
  Phys. Chem. A} \textbf{\bibinfo{volume}{102}}, \bibinfo{pages}{5111}
  (\bibinfo{year}{1998}).

\bibitem[{\citenamefont{Wei et~al.}(2001)\citenamefont{Wei, Guo, and
  Salahub}}]{Salahub01}
\bibinfo{author}{\bibfnamefont{D.}~\bibnamefont{Wei}},
  \bibinfo{author}{\bibfnamefont{H.}~\bibnamefont{Guo}}, \bibnamefont{and}
  \bibinfo{author}{\bibfnamefont{D.}~\bibnamefont{Salahub}},
  \bibinfo{journal}{Phys. Rev. E} \textbf{\bibinfo{volume}{64}}
  (\bibinfo{year}{2001}).

\bibitem[{\citenamefont{Woutersen et~al.}(2001)\citenamefont{Woutersen, Mu,
  Stock, and Hamm}}]{Woutersen01_1}
\bibinfo{author}{\bibfnamefont{S.}~\bibnamefont{Woutersen}},
  \bibinfo{author}{\bibfnamefont{Y.}~\bibnamefont{Mu}},
  \bibinfo{author}{\bibfnamefont{G.}~\bibnamefont{Stock}}, \bibnamefont{and}
  \bibinfo{author}{\bibfnamefont{P.}~\bibnamefont{Hamm}},
  \bibinfo{journal}{Chem. Phys.} \textbf{\bibinfo{volume}{266}}
  (\bibinfo{year}{2001}).

\bibitem[{\citenamefont{Woutersen et~al.}(2002)\citenamefont{Woutersen,
  Pfister, Mu, Kosov, and Stock}}]{Woutersen02}
\bibinfo{author}{\bibfnamefont{S.}~\bibnamefont{Woutersen}},
  \bibinfo{author}{\bibfnamefont{R.}~\bibnamefont{Pfister}},
  \bibinfo{author}{\bibfnamefont{Y.}~\bibnamefont{Mu}},
  \bibinfo{author}{\bibfnamefont{D.}~\bibnamefont{Kosov}}, \bibnamefont{and}
  \bibinfo{author}{\bibfnamefont{G.}~\bibnamefont{Stock}}, \bibinfo{journal}{J.
  Chem. Phys.} \textbf{\bibinfo{volume}{117}}, \bibinfo{pages}{6833}
  (\bibinfo{year}{2002}).

\bibitem[{\citenamefont{Mu and Stock}(2002)}]{Stock02}
\bibinfo{author}{\bibfnamefont{Y.}~\bibnamefont{Mu}} \bibnamefont{and}
  \bibinfo{author}{\bibfnamefont{G.}~\bibnamefont{Stock}}, \bibinfo{journal}{J.
  Phys. Chem. B.} \textbf{\bibinfo{volume}{106}}, \bibinfo{pages}{5294}
  (\bibinfo{year}{2002}).

\bibitem[{\citenamefont{Mu et~al.}(2003)\citenamefont{Mu, Kosov, and
  Stock}}]{Stock03}
\bibinfo{author}{\bibfnamefont{Y.}~\bibnamefont{Mu}},
  \bibinfo{author}{\bibfnamefont{D.}~\bibnamefont{Kosov}}, \bibnamefont{and}
  \bibinfo{author}{\bibfnamefont{G.}~\bibnamefont{Stock}}, \bibinfo{journal}{J.
  Phys. Chem. B.} \textbf{\bibinfo{volume}{107}}, \bibinfo{pages}{5064}
  (\bibinfo{year}{2003}).

\bibitem[{\citenamefont{Nguyen and Stock}(2003)}]{Stock03_1}
\bibinfo{author}{\bibfnamefont{P.}~\bibnamefont{Nguyen}} \bibnamefont{and}
  \bibinfo{author}{\bibfnamefont{G.}~\bibnamefont{Stock}}, \bibinfo{journal}{J.
  Chem. Phys.} \textbf{\bibinfo{volume}{119}}, \bibinfo{pages}{11350}
  (\bibinfo{year}{2003}).

\bibitem[{\citenamefont{Torii and Tasumi}(1998)}]{Torii98}
\bibinfo{author}{\bibfnamefont{H.}~\bibnamefont{Torii}} \bibnamefont{and}
  \bibinfo{author}{\bibfnamefont{M.}~\bibnamefont{Tasumi}},
  \bibinfo{journal}{Journ. of Ram. Spect.} \textbf{\bibinfo{volume}{29}},
  \bibinfo{pages}{81} (\bibinfo{year}{1998}).

\bibitem[{\citenamefont{Schweitzer-Stenner
  et~al.}(2001)\citenamefont{Schweitzer-Stenner, Eker, Huang, and
  Griebenow}}]{Stenner01}
\bibinfo{author}{\bibfnamefont{R.}~\bibnamefont{Schweitzer-Stenner}},
  \bibinfo{author}{\bibfnamefont{F.}~\bibnamefont{Eker}},
  \bibinfo{author}{\bibfnamefont{Q.}~\bibnamefont{Huang}}, \bibnamefont{and}
  \bibinfo{author}{\bibfnamefont{K.}~\bibnamefont{Griebenow}},
  \bibinfo{journal}{J. Am. Chem. Soc.} \textbf{\bibinfo{volume}{123}}
  (\bibinfo{year}{2001}).

\bibitem[{\citenamefont{Levy and Becker}(2001)}]{Levy}
\bibinfo{author}{\bibfnamefont{Y.}~\bibnamefont{Levy}} \bibnamefont{and}
  \bibinfo{author}{\bibfnamefont{O.}~\bibnamefont{Becker}},
  \bibinfo{journal}{J. Chem. Phys.} \textbf{\bibinfo{volume}{114}},
  \bibinfo{pages}{993} (\bibinfo{year}{2001}).

\bibitem[{\citenamefont{Shi et~al.}(2002)\citenamefont{Shi, Olson, Rose,
  Baldwin, and Kallenbach}}]{Shi02}
\bibinfo{author}{\bibfnamefont{Z.}~\bibnamefont{Shi}},
  \bibinfo{author}{\bibfnamefont{C.}~\bibnamefont{Olson}},
  \bibinfo{author}{\bibfnamefont{G.}~\bibnamefont{Rose}},
  \bibinfo{author}{\bibfnamefont{R.}~\bibnamefont{Baldwin}}, \bibnamefont{and}
  \bibinfo{author}{\bibfnamefont{N.}~\bibnamefont{Kallenbach}},
  \bibinfo{journal}{PNAS} \textbf{\bibinfo{volume}{99}}, \bibinfo{pages}{9190}
  (\bibinfo{year}{2002}).

\bibitem[{\citenamefont{Garcia}(2004)}]{Garcia03}
\bibinfo{author}{\bibfnamefont{A.}~\bibnamefont{Garcia}},
  \bibinfo{journal}{Polymer} \textbf{\bibinfo{volume}{45}},
  \bibinfo{pages}{669} (\bibinfo{year}{2004}).

\bibitem[{\citenamefont{Yakubovitch
  et~al.}({\natexlab{a}})\citenamefont{Yakubovitch, Solov'yov, Solov'yov, and
  Greiner}}]{twisting_preprint}
\bibinfo{author}{\bibfnamefont{A.}~\bibnamefont{Yakubovitch}},
  \bibinfo{author}{\bibfnamefont{I.}~\bibnamefont{Solov'yov}},
  \bibinfo{author}{\bibfnamefont{A.}~\bibnamefont{Solov'yov}},
  \bibnamefont{and} \bibinfo{author}{\bibfnamefont{W.}~\bibnamefont{Greiner}},
  \eprint{arXiv: physics/0406093}.

\bibitem[{\citenamefont{Yakubovitch
  et~al.}({\natexlab{b}})\citenamefont{Yakubovitch, Solov'yov, Solov'yov, and
  Greiner}}]{fission_preprint}
\bibinfo{author}{\bibfnamefont{A.}~\bibnamefont{Yakubovitch}},
  \bibinfo{author}{\bibfnamefont{I.}~\bibnamefont{Solov'yov}},
  \bibinfo{author}{\bibfnamefont{A.}~\bibnamefont{Solov'yov}},
  \bibnamefont{and} \bibinfo{author}{\bibfnamefont{W.}~\bibnamefont{Greiner}},
  \eprint{arXiv: physics/0406094}.

\bibitem[{\citenamefont{Guet et~al.}(2001)\citenamefont{Guet, Hobza,
  Spiegelman, and David}}]{LesHouches}
\bibinfo{editor}{\bibfnamefont{C.}~\bibnamefont{Guet}},
  \bibinfo{editor}{\bibfnamefont{P.}~\bibnamefont{Hobza}},
  \bibinfo{editor}{\bibfnamefont{F.}~\bibnamefont{Spiegelman}},
  \bibnamefont{and} \bibinfo{editor}{\bibfnamefont{F.}~\bibnamefont{David}},
  eds., \emph{\bibinfo{title}{Atomic Clusters and Nanoparticles, NATO Advanced
  Study Institute, les Houches Session LXXIII, les Houches, 2000}}
  (\bibinfo{publisher}{EDP Sciences and Springer Verlag, Berlin},
  \bibinfo{year}{2001}).

\bibitem[{\citenamefont{Solov'yov and Connerade}(2004)}]{ISACC2003}
\bibinfo{editor}{\bibfnamefont{A.}~\bibnamefont{Solov'yov}} \bibnamefont{and}
  \bibinfo{editor}{\bibfnamefont{J.-P.} \bibnamefont{Connerade}}, eds.,
  \emph{\bibinfo{title}{Latest Advances in Atomic Cluster Collisions Fission,
  Fusion, Electron, Ion and Photon Impact}} (\bibinfo{publisher}{World
  Scientific Press}, \bibinfo{year}{2004}).

\bibitem[{\citenamefont{Lindgren and Morrison}(1986)}]{Lindgren}
\bibinfo{author}{\bibfnamefont{L.}~\bibnamefont{Lindgren}} \bibnamefont{and}
  \bibinfo{author}{\bibfnamefont{J.}~\bibnamefont{Morrison}},
  \emph{\bibinfo{title}{Atomic Many-Body Theory}}
  (\bibinfo{publisher}{Springer-Verlag, New York, Heidelberg, Berlin},
  \bibinfo{year}{1986}).

\bibitem[{\citenamefont{Hohenberg and Kohn}(1964)}]{Hohenberg64}
\bibinfo{author}{\bibfnamefont{P.}~\bibnamefont{Hohenberg}} \bibnamefont{and}
  \bibinfo{author}{\bibfnamefont{W.}~\bibnamefont{Kohn}},
  \bibinfo{journal}{Phys. Rev.} \textbf{\bibinfo{volume}{136}},
  \bibinfo{pages}{B864} (\bibinfo{year}{1964}).

\bibitem[{\citenamefont{Becke}(1988)}]{Becke88}
\bibinfo{author}{\bibfnamefont{A.}~\bibnamefont{Becke}},
  \bibinfo{journal}{Phys.Rev. A} \textbf{\bibinfo{volume}{38}},
  \bibinfo{pages}{3098} (\bibinfo{year}{1988}).

\bibitem[{\citenamefont{Lee et~al.}(1988)\citenamefont{Lee, Yang, and
  Parr}}]{LYP}
\bibinfo{author}{\bibfnamefont{C.}~\bibnamefont{Lee}},
  \bibinfo{author}{\bibfnamefont{W.}~\bibnamefont{Yang}}, \bibnamefont{and}
  \bibinfo{author}{\bibfnamefont{R.}~\bibnamefont{Parr}},
  \bibinfo{journal}{Phys. Rev. B} \textbf{\bibinfo{volume}{37}},
  \bibinfo{pages}{785} (\bibinfo{year}{1988}).

\bibitem[{\citenamefont{Parr and Yang}(1989)}]{Parr-book}
\bibinfo{author}{\bibfnamefont{R.}~\bibnamefont{Parr}} \bibnamefont{and}
  \bibinfo{author}{\bibfnamefont{W.}~\bibnamefont{Yang}},
  \emph{\bibinfo{title}{Density-Functional Theory of Atoms and Molecules}}
  (\bibinfo{publisher}{Oxford University Press, Oxford, New York},
  \bibinfo{year}{1989}).

\bibitem[{\citenamefont{Bax}(2003)}]{Bax03}
\bibinfo{author}{\bibfnamefont{A.}~\bibnamefont{Bax}}, \bibinfo{journal}{Prot.
  Sci.} \textbf{\bibinfo{volume}{12}}, \bibinfo{pages}{1}
  (\bibinfo{year}{2003}).

\bibitem[{\citenamefont{Sheik et~al.}(2002)\citenamefont{Sheik, Sundararajan,
  Hussain, and Sekar}}]{RamachandranPlot}
\bibinfo{author}{\bibfnamefont{S.}~\bibnamefont{Sheik}},
  \bibinfo{author}{\bibfnamefont{P.}~\bibnamefont{Sundararajan}},
  \bibinfo{author}{\bibfnamefont{A.}~\bibnamefont{Hussain}}, \bibnamefont{and}
  \bibinfo{author}{\bibfnamefont{K.}~\bibnamefont{Sekar}},
  \bibinfo{journal}{Bioinformatics} \textbf{\bibinfo{volume}{18}},
  \bibinfo{pages}{1548} (\bibinfo{year}{2002}).

\bibitem[{\citenamefont{Voet and Voet}(2004)}]{Biochemistry_book}
\bibinfo{author}{\bibfnamefont{D.}~\bibnamefont{Voet}} \bibnamefont{and}
  \bibinfo{author}{\bibfnamefont{J.}~\bibnamefont{Voet}},
  \emph{\bibinfo{title}{Biochemistry}} (\bibinfo{publisher}{John Willey and
  Sons, Inc., USA}, \bibinfo{year}{2004}).

\end{thebibliography}
\end{document}